\documentstyle[12pt,aasms4]{article}

\lefthead{Williams et al.}
\righthead{Optical Afterglow from GRB 971227}

\begin{document}

\title{LOTIS Search for Early Time Optical Afterglows:  GRB~971227}

\author{G.~G.~Williams\altaffilmark{1},
	H.~S.~Park\altaffilmark{2},
	E.~Ables\altaffilmark{2},
	D.~L.~Band\altaffilmark{5},
	S.~D.~Barthelmy\altaffilmark{3,8},
	R.~Bionta\altaffilmark{2},
	P.~S.~Butterworth\altaffilmark{3,9},
	T.~L.~Cline\altaffilmark{3},
	D.~H.~Ferguson\altaffilmark{6},
	G.~J.~Fishman\altaffilmark{4},
	N.~Gehrels\altaffilmark{3},
	D.~H.~Hartmann\altaffilmark{1},
	K.~Hurley\altaffilmark{7},
	C.~Kouveliotou\altaffilmark{4,8},
	C.~A.~Meegan\altaffilmark{4},
	L.~Ott\altaffilmark{2},
	E.~Parker\altaffilmark{2} and
	R.~Porrata\altaffilmark{2}}


\altaffiltext{1}{Dept. of Physics and Astronomy, Clemson University, Clemson, SC 29634--1911}
\altaffiltext{2}{Lawrence Livermore National Laboratory, Livermore, CA 94550}
\altaffiltext{3}{NASA/Goddard Space Flight Center, Greenbelt, MD 20771}
\altaffiltext{4}{NASA/Marshall Space Flight Center, Huntsville, AL 35812}
\altaffiltext{5}{CASS 0424, University of California, San Diego, La Jolla, CA 92093}
\altaffiltext{6}{Dept. of Physics, California State University at Hayward, Hayward, CA 94542}
\altaffiltext{7}{Space Sciences Laboratory, University of California, Berkeley, CA 94720--7450}
\altaffiltext{8}{Universities Space Research Association, Columbia, MD 21044--3498}
\altaffiltext{9}{Raytheon Systems, Lanham, MD 20706}


\begin{abstract}
We report on the very early time search for an optical afterglow  
from GRB~971227 with the Livermore Optical Transient
Imaging System (LOTIS).  LOTIS began imaging the `Original' 
BATSE error box of GRB~971227 $\sim 14$~s after the
onset of gamma-ray emission.  Continuous monitoring
of the position throughout the evening 
yielded a total of 499 images (10~s integration).  
Analysis of these images revealed no steady optical afterglow 
brighter than R=$12.3\pm0.2$ in any single image.
Coaddition of the LOTIS images also failed to uncover 
transient optical emission.  In particular, assuming 
a constant early time flux, no optical afterglow 
brighter than R=$14.2\pm0.2$ was present within the 
first 1200~s and no optical afterglow brighter than 
R=$15.0\pm0.2$ was present in the first 6.0 h. 

Follow up observations by other groups revealed 
a likely X-ray afterglow and 
a possible optical afterglow.  Although
subsequent deeper observations could not
confirm a fading source, we show that
these transients are not inconsistent with our 
present knowledge of the characteristics 
of GRB afterglows.  We also demonstrate that
with the upgraded thermoelectrically cooled CCDs,
LOTIS is capable of either detecting very
early time optical afterglow or placing 
stringent constraints on the relationship 
between the gamma-ray emission and the 
longer wavelength afterglow in relativistic 
blast wave models.
\end{abstract}




%

\section{Introduction}

Recent observations of long wavelength (X-ray, optical 
and radio) afterglows of gamma-ray bursts (GRBs) have
provided precise localizations 
(e.g., \cite{levine98}; \cite{vanparadijs97}; \cite{frail97}). 
Redshift measurements of the apparent host galaxies 
strongly suggest that GRBs are located at very large distances 
(\cite{metzger97}; \cite{kulkarni98}; \cite{djorgovski98}).
These distances together with the measured 
gamma-ray fluxes imply that GRBs are extremely 
energetic phenomena (e.g., \cite{kulkarni98}).  X-ray and 
optical afterglows have consistently displayed 
a power law decay in time 
(\cite{wijers97}; \cite{groot98}; \cite{bloom98})
with some notable deviations (\cite{galama97a}).  
Flux variations in radio afterglows have been
interpreted as dif\/fractive scintillation ef\/fects.  
The decreasing amplitude of these 
variations with time suggests a relativistically 
expanding source (\cite{goodman97}; \cite{frail97}).

All of the above GRB and afterglow characteristics 
are, for the most part, consistent with a 
relativistic blast wave model (\cite{rees92}; \cite{meszaros93}). 
However it is still unclear what physical 
mechanism is producing the underlying energetic 
fireballs.  In fact a definitive signature of 
the central engine may be impossible to 
observe directly due to the optically thick 
baryon-loaded $e^{\pm}, \gamma$ fireball (\cite{goodman86}).

Perhaps more can be learned about the source 
of GRBs through early time observations 
of the afterglow.  These observations may 
reveal whether the central engine deposits 
energy impulsively (\cite{meszaros93}) or in an 
unsteady wind (\cite{meszaros94}; \cite{paczynski94}). 
The impulsive blast wave model produces both the 
GRB and the afterglow in an external shock. The 
early time afterglow flux is predicted to roughly 
scale to the initial gamma-ray flux.  At later times 
the flux decays in a power law fashion as the 
shock slows due to the sweeping up of the 
external circum-burster material (CBM).

In the unsteady wind model the GRB is produced by 
collisions between internal shocks 
(e.g., \cite{sari97}) while the afterglow is
produced in the external shock.  Both models 
predict similar late time long wavelength 
behavior. However the early time optical 
afterglow in the case of the unsteady wind
may be more complex and is less constrained 
by the initial gamma-ray flux.

Observations of GRB~990123 have established the 
existence of prompt optical emission (\cite{akerlof99}).
The peak optical flux appears to be slightly 
delayed relative to the peak gamma-ray flux.
Sari \& Piran (1999) have suggested that this
delay is indicative of the dif\/ferent production 
regions for the gamma-ray and the prompt optical 
emission.  The GRB may form in the internal shocks, 
while the prompt optical and X-ray emission result 
from the reverse and forward external shocks, 
respectively (e.g., \cite{sari99}).  Previously 
reported LOTIS upper limits (\cite{park97a}; 
\cite{williams97}; \cite{park98a}), 
as well as those presented here and in the
future may provide constraints on the emission 
sites in the expanding GRB fireball.

\section{Observations Following GRB~971227}

LOTIS is a robotic wide field-of-view telescope 
capable of responding to a GRB Coordinates 
Network (GCN; \cite{barthelmy97}) trigger within 
10~s of the start of a burst.  The system 
consists of four cameras mounted in a 
$2 \times 2$ array which cover a total 
field-of-view of $17.\!\!^{\circ}4 \times 17.\!\!^{\circ}4$.
To ensure maximum light gathering power the 
cameras have not been equipped with standard 
filters.  Since the response of the 2k~$\times$~2k
Loral CCDs most closely resembles the R-band, we 
assume R-band photometry.  The small errors
introduced by this assumption have been accounted 
for in the analysis and are included in the stated
uncertainties.  Specific information 
on the LOTIS system and operational 
method is provided in Park et al.\ (1997a)\nocite{park97a},
Park et al.\ (1998b)\nocite{park98b} and at the LOTIS 
homepage\footnote[1]{http://hubcap.clemson.edu/$\sim$ggwilli/LOTIS/}.

GRB~971227 was detected by the Burst and 
Transient Source Experiment (BATSE) instrument 
aboard Compton Gamma-Ray Observatory (CGRO) on 
December~27.34938 (8:23:06.72 UT).  This burst was 
simultaneously detected by both the Gamma-Ray Burst 
Monitor (GRBM) and the Wide Field Camera WFC2 aboard 
the BeppoSAX satellite (\cite{coletta97}).  GRB~971227 
consisted of a single main peak and a weaker second 
peak with a total duration of $\sim 7$~s
(T$_{90} = 6.784$; \cite{meegan98}).  
BATSE measured a peak photon flux and fluence 
of $3.3\pm0.2$ cm$^{-2}$ s$^{-1}$ 
(64 ms; 50--300 keV) and
$9.3\pm1.4\times10^{-7}$ erg cm$^{-2}$ ($>$ 25 keV)
respectively (\cite{woods97}).  The peak flux measured 
by the BeppoSAX WFC2 was 1.8 Crab 
($3.7\times10^{-8}$ erg cm$^{-2}$ s$^{-1}$;
2--10 keV; \cite{coletta97}).  The `Original' BATSE coordinates 
for the burst were RA=$13^{\rm h}10^{\rm m}07^{\rm s}$, 
Dec=$+65^{\circ}48^\prime$ (J2000.0).  These `Original' 
coordinates are calculated using the first 1~or~2~seconds 
of the burst light curve, sacrificing positional accuracy 
in favor of rapid dissemination (\cite{barthelmy97}
and the GCN homepage\footnote[2]{http://gcn.gsfc.nasa.gov/gcn}).
The initial BeppoSAX WFC2 position was 
RA=$12^{\rm h}57^{\rm m}29^{\rm s}$, 
Dec=$+59^{\circ}16.\!\!^{\prime}3$ (J2000.0) with an 
error radius of $10^\prime$ (\cite{coletta97}).
GRB~971227 was also detected by the Ulysses gamma-ray 
burst monitor (\cite{hurley92}).  
The Ulysses/BATSE triangulation provided a 
$\sim 7^{\prime}$ wide annulus intersecting the 
final 3$\sigma$ BATSE and the initial BeppoSAX WFC2 
error circles.

Follow up observations 14~h after the GRB
by the BeppoSAX Narrow Field Instruments (NFI) 
detected two previously unknown X-ray sources 
within the initial WFC2 error box.  One source 
(1SAX~J1257.5+5915) did not show significant 
variation during the 20 h observation period.  
The second source (1SAX~J1257.3+5924), with
coordinates RA=$12^{\rm h}57^{\rm m}33^{\rm s}$ ($194.3875^{\circ}$), 
Dec=$59^{\circ}15^\prime 27^{\prime\prime}$ ($59.2575^{\circ}$; J2000.0)
and an error radius of $1^{\prime}\!.5$,
had a time average flux of 
$3\times10^{-13}$ erg cm$^{-2}$ s$^{-1}$ (2--10~keV)
during the first 8~h but was undetectable at the 
3$\sigma$ level for the remainder of the 
observation (\cite{piro97}).  
Although the association of the fading X-ray source 
with GRB~971227 could not be confirmed it is considered
the most likely candidate for the X-ray afterglow.

\begin{figure}
\plotone{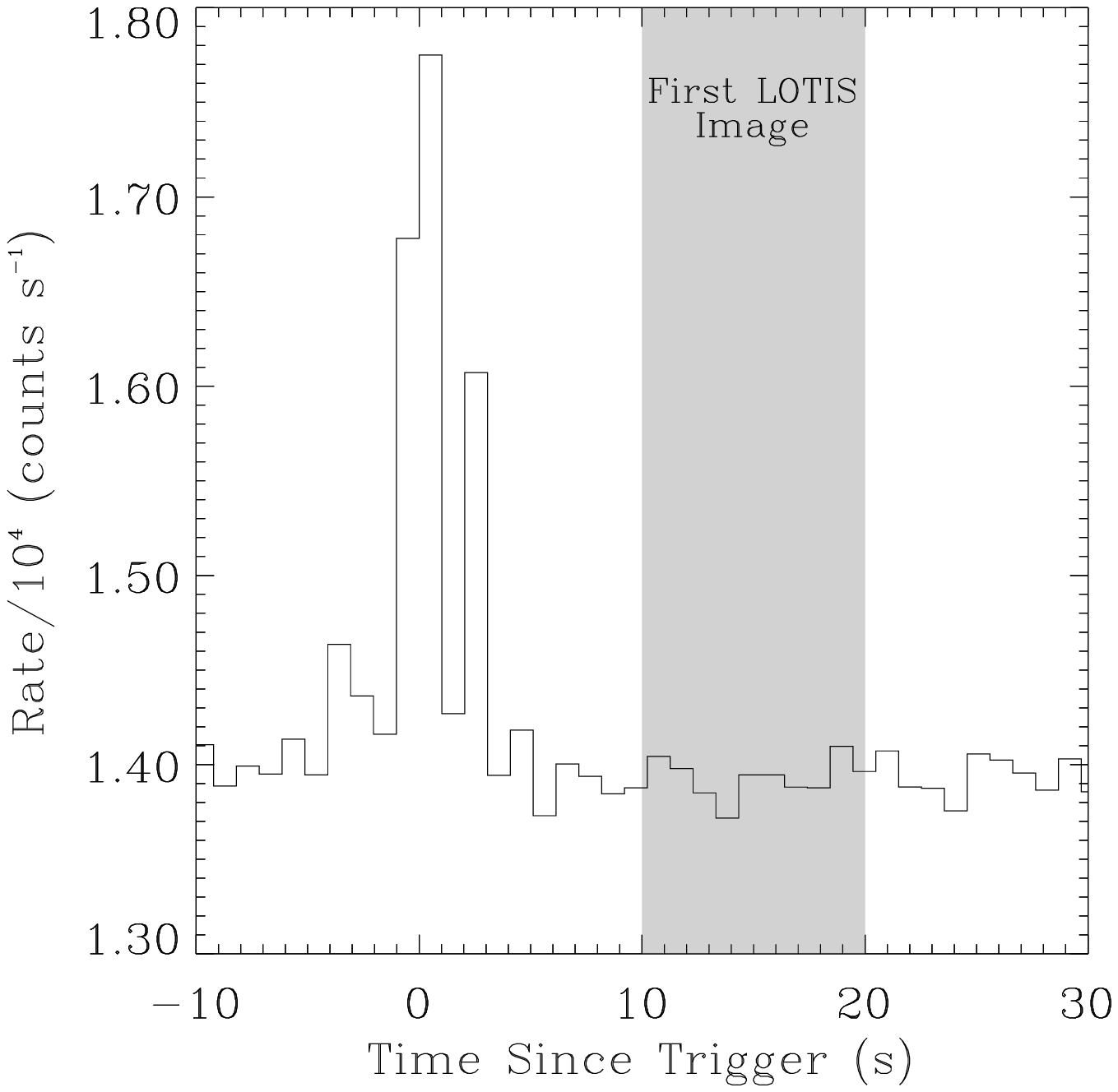}
\caption{BATSE Trigger 6546 (GRB 971227) light curve (E $> 20$ keV).
The shaded area represents the integration time of the first
LOTIS image. \label{fig:grb6546lc}}
\end{figure}

\begin{figure}
\plotone{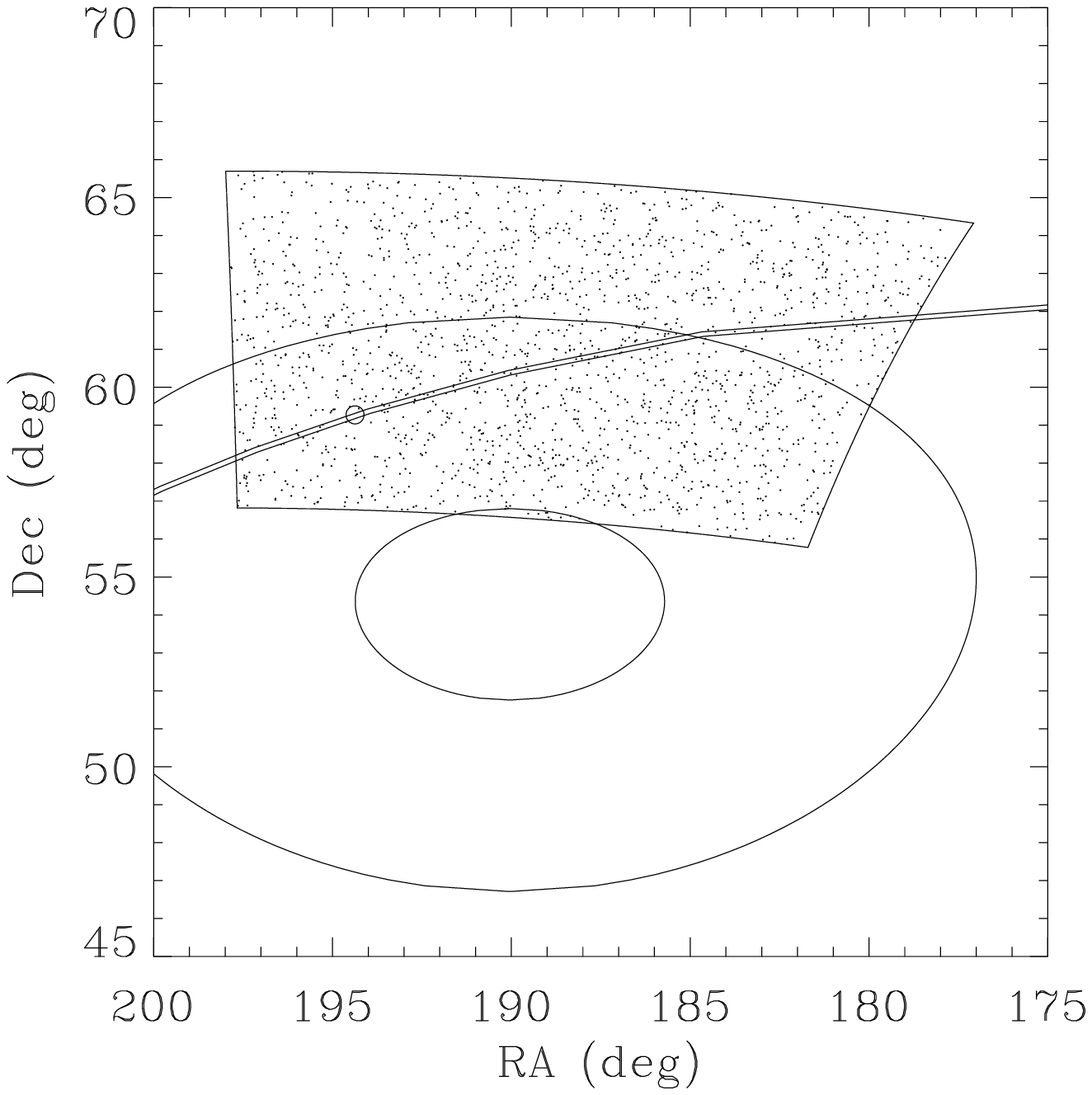}
\caption{LOTIS camera 1 coverage plot for GRB 971227.  Each 
of the dots represents a stellar object detected by the 
system.  The large ellipses represent the final 
1$\sigma$ and 3$\sigma$ BATSE error boxes.  The arc is the
3$\sigma$ IPN annulus and the small circle is the BeppoSAX WFC2
error circle. \label{fig:cov6546}}
\end{figure}

LOTIS received the GCN coordinates at 8:23:10.9~UT and after 
a slew time of 5.8~s began imaging the area near the `Original' 
GRB coordinates.  The first LOTIS image began 10.0~s 
after the BATSE trigger.  Figure~\ref{fig:grb6546lc} 
shows the BATSE gamma-ray (E~$>$~20~keV) light 
curve for GRB~971227 (\cite{meegan98}).  
The shaded area represents the duration 
(10~s integration) of the first LOTIS image.  
Although the `Original' BATSE coordinates 
were $\sim 6.\!\!^{\circ}7$ from 
the more precise BeppoSAX WFC2 coordinates, the 
LOTIS wide field-of-view camera system allowed 
full coverage of the X-ray counterpart error box.
Figure~\ref{fig:cov6546} shows the area covered 
by LOTIS camera~1. Coverage of cameras 2, 3 and 4 is
not shown.  The center of the $2 \times 2$ array of 
cameras is in the upper left hand corner.
Each of the dots represents a stellar
object detected (3$\sigma$) by the system.
The final 1$\sigma$ and 3$\sigma$ BATSE error circles
as well as the Interplanetary Network (IPN; \cite{hurley98}) annulus 
and BeppoSAX WFC2 error circles are also shown. 





\section{Data Analysis and Results}
In the 6 dark hours following GRB~971227 LOTIS 
obtained 499 images (10~s integration) which 
fully covered the X-ray counterpart error box.  
All the images were dark subtracted and flat fielded. 
Visual inspection was carried 
out on the full images ($8.\!\!^{\circ}8 \times 8.\!\!^{\circ}8$)
as well as on a subsection of the images
($1.\!\!^{\circ}1 \times 1.\!\!^{\circ}1$)
centered on the suspected X-ray counterpart.   
No astrophysical transient events consistent with 
the point spread function of the system were found.  
DAOPHOT (\cite{stetson87}) routines were used to identify 
all stellar objects brighter than 3$\sigma$ above 
the background for the entire LOTIS data set.  
All objects were traced from image to image and 
variable objects were flagged.  These objects 
were then carefully inspected by eye and in all 
cases discarded as a possible afterglow candidate.  
All objects found within the initial 
10$^\prime$ radius X-ray counterpart 
error circle were identified with known stars in
the Guide Star Catalog (GSC; \cite{jenkner90}).
No transient events or previously unidentified
sources brighter than R=$12.3\pm0.2$ were 
found within the error circle. 
 
Since the simplest fireball model predicts a 
constant early-time optical flux we coadded LOTIS
images to increase the signal-to-noise ratio.
We first coadded and analyzed every set of ten images 
(1--10, 11--20, 21--30,...,491--499).  Including the 
10~s CCD readout time each set of images
corresponds to an interval of $\sim 200$~s.  In this manner 
the LOTIS sensitivity to short term transient 
events was increased to a limiting magnitude of 
R=$13.2\pm0.2$.  No previously unknown sources were 
detected.  We then coadded consecutive sets of images 
in increments of ten (1--10, 1--20, 1--30,...,1--499). 
All stellar objects in the summed images were matched
with previously identified objects in the USNO-A1.0 
Catalog (\cite{monet96}).  No new sources brighter 
than R=$14.2\pm0.2$ were revealed during the 
first 1200~s (1--56) and no new sources brighter than 
R=$15.0\pm0.2$ were revealed during the first
21300~s (1--490).  We present the results of summing 
the first 1200~s of data because LOTIS was 
in `burst mode' during that time.  While in  
`burst mode' LOTIS obtains one 10~s image every
20~s.  At later times LOTIS was in `sky patrol mode' 
obtaining one 10~s image every 40~s.

\begin{figure}
\plotone{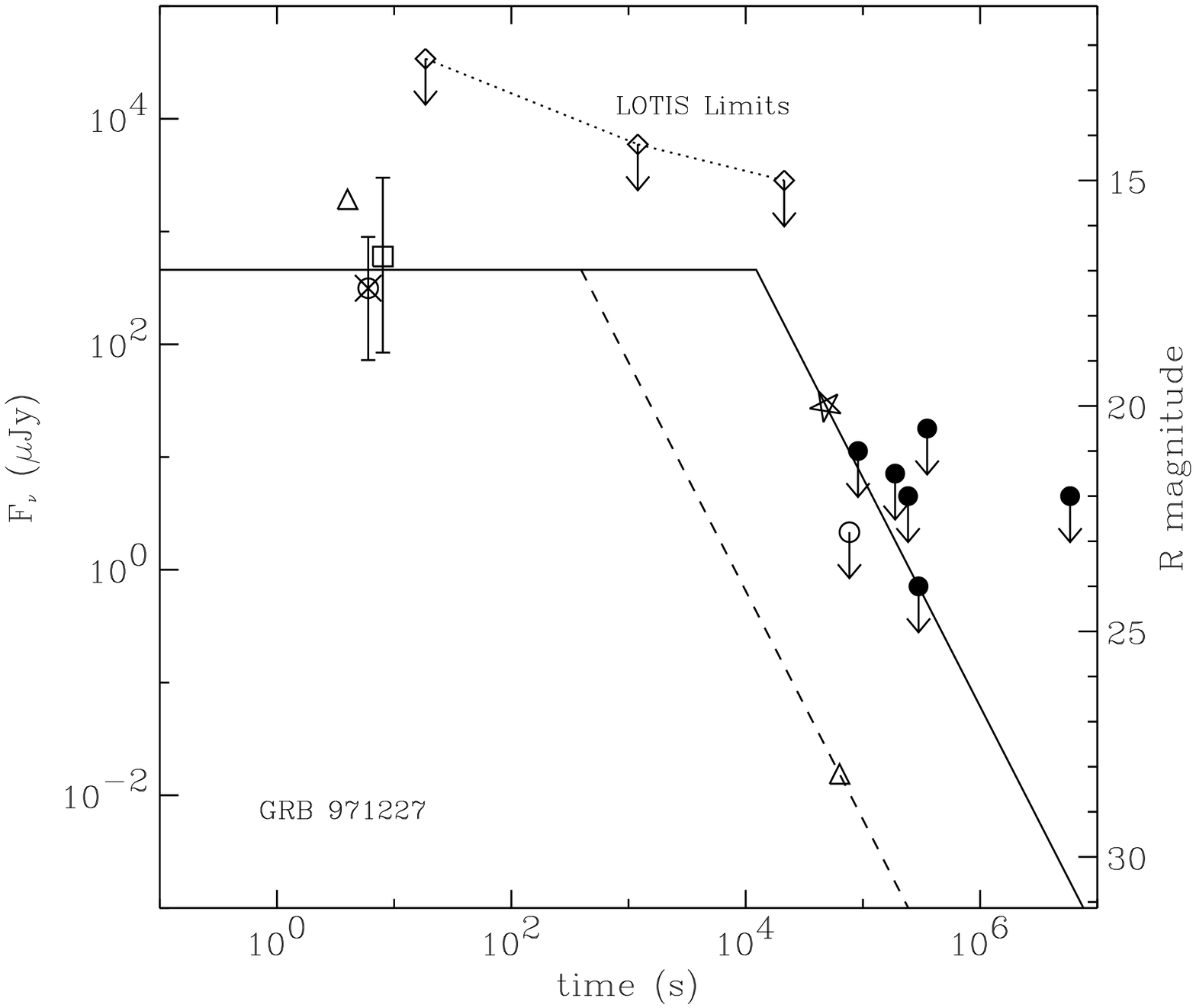}
\caption{Broadband light curve of GRB 971227.  The diamonds
are the LOTIS upper limits.  The crossed circle and square
are the gamma-ray (100~keV) and X-ray (5~keV) flux densities 
derived from BATSE data.  The triangles are X-ray flux densities
derived from BeppoSAX data.  The star represents the possible 
optical afterglow.  R-band upper limits are plotted as filled 
circles.  The open circle is an upper limit which did not cover the
area of the suspected optical afterglow.
The solid line illustrates the predictions
from a simple fireball model.  \label{fig:opt1227}}
\end{figure}

Figure~\ref{fig:opt1227} shows a broadband light curve 
of the afterglow of GRB~971227 based on the simple fireball 
model (\cite{meszaros93}; \cite{wijers97}).  The data for this
plot are listed in Table~\ref{tbl:flux}.  The crossed circle and 
the square are the gamma-ray (100 keV) and 
X-ray (5~keV) flux densities derived from BATSE data.  
We fit the High Energy Resolution Burst data (HERB; consisting of 
128 separate high energy resolution spectra from four Large Area
Detectors) from 0.033~s to 6.976~s (29.16--1881.5 keV)
to the Band functional form 
(\cite{band93}) and extracted the value of the
fit at 100~keV and at 5~keV (extrapolation).  
The error bars reflect the range of the best fit Band
functional form.  The times have been of\/fset by
$\pm 1$ s to avoid confusion. 

The BeppoSAX data are represented by triangles.
The first triangle is the approximate X-ray flux 
density ($F_{peak:{\rm{2-10 \, keV}}}$/8 keV) 
measured by the WFC2 during GRB~971227.  The second 
triangle is the approximate flux density ($F_{\rm{2-10 \, keV}}$/8 keV) 
of the suspected X-ray afterglow measured with 
the BeppoSAX NFI.

The LOTIS upper limits are plotted as diamonds.  
The first point indicates the upper limit of 
R=$12.3\pm0.2$ derived from the first LOTIS image.
The second and third diamonds are the upper limits of 
R=$14.2\pm0.2$ and R=$15.0\pm0.2$ achieved after 
adding the first 1200~s and 21300~s of LOTIS data.  
The dotted straight line connecting the LOTIS points 
is an approximation of how the sensitivity changes 
with integration time.  The line changes slope because
of the decrease in total integration time per unit real 
time when LOTIS changes from `burst mode' to `sky patrol mode'. 

The circles plotted at late times are the R-band upper 
limits obtained by several dif\/ferent groups.  The 
photometric values have not been dereddened since 
the burst occurred at high galactic latitude (l=121.464,b=57.853).
The single open circle is the upper limit quoted 
by \nocite{groot97} Groot et al. (1997).  The area 
covered in this observation was the inner 
$12^\prime \times 12^\prime$ of the initial X-ray 
counterpart error box (\cite{groot97}).  It is 
important to note that this observation 
does not cover the area of the optical afterglow 
candidate suggested by \nocite{castro97} 
Castro-Tirado et al. (1997) which is plotted as a star.
Preliminary photometry of the afterglow candidate 
yielded an apparent magnitude of R=$19.5$ (\cite{castro97}).
Further analysis found an apparent magnitude of 
R=$20.0 \pm 0.3$ (\cite{bartolini98}). 

The solid line is a fit to the data assuming a simple 
fireball model.  In this case
the synchrotron break frequency evolves as 
${\nu}_m \propto t^{-3/2}$ due to the decay of the 
bulk Lorentz factor as the blast wave engulfs
increasing amounts of CBM.  Prior to passage of the 
synchrotron break frequency through the observation
band the early time flux varies as $F_{{\nu}_m} \propto t^0$.
At later times, after the break frequency passes through 
the observation band, the flux decays as
$F_{ob} \propto F_{{\nu}_m}({\nu}_{ob}/{\nu}_m)^{{\beta}^\prime}$
or $F_{ob} \propto t^{(3/2){\beta}^\prime} \equiv t^{-\alpha}$ 
where ${\beta}^{\prime}$ is the high energy spectral index
(\cite{meszaros93}; \cite{wijers97}).  We adjust the predicted
early time flux to the initial gamma-ray and X-ray fluxes. 
The optical decay is fit with a power law through the 
possible afterglow detection and the most constraining 
late time upper limit of \nocite{djorgovski97} 
Djorgovski et al. (1997). Using this method we find a 
lower limit on the power law decay index of $\alpha > 2.03$.  
The optical counterparts detected to date have decay 
indices ranging from approximately $\alpha = 1.17$ for GRB~980703 
(\cite{bloom98}) to $\alpha = 2.1$ for GRB~980326 (\cite{groot98}). 
If the transient detected by \nocite{castro97} 
Castro-Tirado et al. (1997) is indeed associated with 
GRB~971227 it displayed one of the fastest decays of the 
optical counterparts detected to date.  

To constrain the relationship between GRB 
properties and the intensity of prompt optical 
emission, we compare GRB~971227 with GRB~990123.
The ratios of the peak flux and fluence
for the two events is
$F_{peak:{\rm{971227}}}/F_{peak:{\rm{990123}}} = 0.20$ and 
$S_{\rm{971227}}/S_{\rm{990123}} = 1.91\times10^{-3}$
(\cite{woods97};\cite{meegan98}).
If the prompt optical flux scales with the gamma-ray 
fluence then GRB~971227 would have had a peak magnitude 
of m $= 15.8$ and therefore would have been undetectable by LOTIS.
If the peak optical flux scales with the peak gamma-ray 
flux then GRB~971227 would have produced an optical 
flash which peaked at m $=10.7$.  If this flash occured 
during the first LOTIS image it would have easily been
detected.  However the level of optical 
emission during the first LOTIS image depends on the 
time of the optical peak and its rate of decay; 
both uncertain parameters.  A full comparison between 
properties of GRB~990123 and the set of GRBs which 
have resulted in LOTIS upper limits will be presented 
in a future publication.

\section{Conclusion}

Although identifications of an X-ray and optical afterglow of
GRB~971227 are inconclusive we have shown that the properties
of these suspected transient events can be made consistent
with the fireball model.  Assuming the transients are associated 
with the burst we find a lower limit on the index of 
the power law decay for the afterglow of GRB~971227 
of $\alpha > 2.0$.

In addition to characterizing 
the early time afterglow from GRBs, early
detection could allow dif\/ferentiation between physical 
processes causing the GRB itself and the afterglow.
Although LOTIS was able to provide meaningful
constraints on early time optical afterglow from
GRB~971227 at intermediate times ($10^{1-4}$~s),
these constraints were not stringent since this 
was not a particularly intense burst in terms of peak flux.
A recent upgrade to thermoelectrically cooled CCDs has 
increased the LOTIS sensitivity by a factor of 10
(R=12.0 to R=14.5 in a single exposure).  With 
this upgrade in place LOTIS is now much more likely
to detect or significantly constrain early time 
optical afterglows.
 
\acknowledgments

We thank Mark Leising for useful comments.  This work 
was supported by the U.S. Department of  Energy, under contract
W-7405-ENG-48 and NASA contract S-57771-F.  Gamma-ray burst 
research at UCSD (D. Band) is supported by the CGRO 
guest investigator program.  K. Hurley acknowledges JPL 
Contract 958056 for Ulysses operations and NASA 
Grant NAG5-3811 for IPN work.

\clearpage

\begin{table*}
\begin{center}
\begin{tabular}{lr@{.}lclc} \tableline\tableline
log $t$ & \multicolumn{2}{c}{$F_{\nu}$} & Band & Instrument & Ref. \\
(s) & \multicolumn{2}{c}{($\mu$Jy)} &  &  &  \\
\tableline
0.6021 &1934& &X &BeppoSAX(WFC2)  &1 \\
0.8454 &314& &$\gamma$ &BATSE  &Band Fit\\
0.8454 &601& &X &BATSE  &Band Fit \\
1.2672 &$<$34144& &R &LOTIS  &2 \\
3.0792 &$<$5933& &R &LOTIS  &- \\
4.3284 &$<$2840& &R &LOTIS  &- \\
4.6852 &28&400 &R &CAHA Tele.  &3,12 \\
4.8006 &0&0155  &X &BeppoSAX(NFI)  &4 \\
4.8847 &$<$2&154 &R &Isaac Newton Tele.  &5 \\
4.9580 &$<$11&306 &R &Palomar 60 in Tele. &6 \\
5.2751 &$<$7&134 &R &Kitt Peak 0.9 m Tele. &7 \\
5.3853 &$<$4&501 &R &Loiano 1.5 m Tele. &8,12 \\
5.4744 &$<$0&713 &R &Keck &9 \\
5.5483 &$<$17&919 &R &DAO 1.8 m Plaskett Refl.  &10 \\
6.7680 &$<$4&501 &R &1.2 m OHP Tele.  &11 \\
\hline
\end{tabular}
\end{center}




\tablenum{1}
\caption{Observed fluxes and upper limits of GRB~971227 and its afterglow. \label{tbl:flux}}

\tablerefs{(1) \cite{coletta97}; (2) \cite{park97a}; (3) \cite{castro97};
(4) \cite{piro97}; (5) \cite{groot97}; (6) \cite{ramaprakash97};
(7) \cite{galama97b}; (8) \cite{bartolini97}; (9) \cite{djorgovski97};
(10) \cite{bond97}; (11) \cite{ilovaisky97}; (12) \cite{bartolini98}}


\end{table*}




\clearpage

\clearpage

%

\end{document}